# Point spread function engineering for spiral phase interferometric scattering microscopy enables robust 3D single-particle tracking

Nathan J. Brooks[1], Chih-Chen Liu[1,2], Chia-Lung Hsieh[1,2]*

[1]Institute of Atomic and Molecular Sciences (IAMS), Academia Sinica, Taipei 10617, Taiwan
[2]Department of Physics, National Taiwan University, Taipei 10617, Taiwan
*Correspondence: clh@gate.sinica.edu.tw

**Abstract**

Interferometric scattering (iSCAT) microscopy is currently among the most powerful techniques available for achieving high-sensitivity single-particle localization. This capability is realized through homodyne detection, where interference with a reference wave offers the promise of exceptionally precise three-dimensional (3D) localization. However, the practical application of iSCAT to 3D tracking has to date been hampered by rapid oscillations in the signal-to-noise ratio (SNR) as particles move along the axial direction. In this study, we introduce a novel strategy based on back pupil plane engineering, wherein we use a spiral phase mask to re-distribute the phase of the scattered field of the particle uniformly across phase space, thus ensuring consistent SNR as the particle moves throughout the focal volume. Our findings demonstrate that this modified spiral phase iSCAT exhibits greatly enhanced localizability characteristics. We substantiate our theoretical results with numerical and experimental demonstrations, showcasing the practical application of this approach for high-precision, ultrahigh-speed (20,000 frames per second) 3D tracking, and characterization of freely diffusing nanoparticles.

## INTRODUCTION

Localization and tracking of single particles have broad applications, including particle tracking velocimetry [1], nanoparticle characterization [2, 3], single-molecule biophysics [4, 5], and super-resolution localization microscopy [6, 7]. The ability to visualize single particles allows for measuring their individual properties and dynamics, which is valuable for characterizing heterogeneous samples. In many instances, the particles of interest are small in size, encompassing synthetic nanoparticles, liposomes for drug delivery, and biological nanoparticles such as virus particles and cell vesicles. Optical microscopes are employed for the observation and high-resolution tracking of these nanoparticles, using a microscope objective to collect the optical signals emitted by individual particles. Conventional microscopes, however, are not ideal for particle localization along the axial direction due to their limited depth of focus, complicating the tracking of nanoparticles in 3D.

To address this challenge, scientists have employed PSF engineering to encode the axial position of the particle into the intensity in the lateral plane. The PSF engineering is typically accomplished by inserting a phase plate at the back pupil plane of the microscope objective, introducing a designed phase modulation in the Fourier plane of the sample. Previous studies have demonstrated various phase patterns for PSF engineering in fluorescence single-particle tracking (SPT), including astigmatism [8], double-helix PSF [9], tetrapod PSFs [10], and optimal PSF for 3D imaging [11], achieving improved localization precision over an extended axial range. While fluorescence provides high contrast for SPT, especially in complex samples, the photobleaching effect fundamentally limits observation time and image acquisition rate. Furthermore, characterizing nonfluorescent nanoparticles is nontrivial because fluorophore labeling of the nanoparticles of interest is not only time-consuming but also raises concerns about introducing labeling artifacts.

The challenges posed by the limited photon budget and labeling artifacts find solutions in the detection of linear scattering signals emitted by particles [12]. Various scattering-based optical techniques have been demonstrated for the imaging and tracking of single nanoparticles. These include interferometric scattering microscopy (iSCAT) [13-15], holographic microscopy [16-19], dark-field microscopy [20-22], surface plasmon resonance microscopy [23, 24], and image scanning microscopy [25]. Differing in their illumination and detection schemes, these techniques offer distinct sensitivities and tracking capabilities. Notably, iSCAT microscopy introduces a coherent reference beam that interferes with the scattered signal in a common-path configuration, providing remarkable measurement stability and detection sensitivity. This makes it feasible to detect very small nanoparticles whose scattering signals would be difficult to capture with dark-field-based approaches [26-28]. Furthermore, interferometric detection provides an additional means for the axial localization of particles beyond what is possible with non-interferometric techniques, with axial information encoded in the interference contrast that is sensitive to the phase of the scattered signal. Using iSCAT, very small nanoparticles, such as virus particles, cell vesicles, and gold nanoparticles, can be directly visualized at the single particle level, and their nanoscopic displacements measured in 3D at a very high speed [29-31]. Moreover, the interferogram-derived complex polarizability of the particle enables the estimation of its refractive index, establishing a direct connection to its density and composition [32].

Despite the success of iSCAT in SPT, it faces a challenge in robust 3D tracking due to flickering interference contrast when the particle displaces along the axial direction. Specifically, the particle contrast largely vanishes at certain axial positions where the phase difference between the signal and the reference is $(2n + 1)\pi/2$, a characteristic manifested in the interferometric PSF (iPSF) of iSCAT microscopy [33]. The low interference contrast in these planes complicates the detection and accurate localization of the particle [34]. Furthermore, as particle contrast is employed for localization in the axial direction, the quantitative determination of the scattering magnitude becomes nontrivial, leading to difficulty in estimating particle density and refractive index.

In this study, we introduce a strategy to unlock the potential of iSCAT imaging for robust 3D SPT by combining the exquisite phase sensitivity of iSCAT with iPSF engineering. In contrast to fluorescence microscopy, where PSF engineering primarily extends the axial tracking range of a particle, we show that its introduction to iSCAT carries a distinct set of advantages. By introducing a spiral phase modulation in the back pupil plane of the microscope objective, the spiral phase iPSF (SP-iPSF) improves upon the standard iPSF in three key ways, all of which significantly benefit the task of 3D SPT: 1) the scattered phase information of the particle is uniformly distributed, leading to a persistent, axially varying iPSF that allows robust fitting at all points throughout the focal volume, 2) symmetry about the focus is broken and similarity between periodic local minima in the fitting process is reduced, allowing accurate assessment of the direction and absolute positioning of the particle, and 3) the axial position is encoded entirely in the shape of the twisting iPSF, allowing the contrast to be decoupled and fit independently for scattering amplitude measurement. We demonstrate that this SP-iPSF enhances localization precision compared to the conventional iPSF of iSCAT. This improvement allows us to measure 3D diffusion trajectories of single nanoparticles with higher accuracy and precision. Furthermore, as the SP-iPSF encodes the axial information in the iPSF shape, the interference contrast becomes an independent variable that directly represents the magnitude of the scattered field. Exploiting this complete decoupling of scattering strength and axial position, we demonstrate that SP-iPSF allows for accurate and precise measurement of the particle polarizability simultaneously.

## RESULTS

**Stabilized particle contrast and localizability throughout the focal volume**

The image formation in iSCAT is expressed by the equation:

$$I = |E_r|^2 + |E_s|^2 + |E_r||E_s|\cos\theta \qquad \text{Eq. (1)}$$

where $E_r$ is the reference wave which is partially reflected from the glass coverslip (approximately a plane wave under widefield illumination), $E_s$ is the scattered wave created by interaction with the particle sample, and $\theta$ denotes the phase difference between the two waves. In the context of scattering from subwavelength particles, the image is described by the iPSF. Using a vectorial diffraction model [33], we perform numerical simulations to depict the iPSF of a particle in a widefield reflection iSCAT microscope. In our simulation, the aberration induced by the refractive index mismatch between the sample and coverslip is considered.

The phase difference $\theta$ incorporates several effects, including the Gouy phase, a spherical phase resulting from displacement from the focus, and aberrations. However, the dominant contributing factor is an additional phase term arising from the axial displacement of the scatterer relative to the coverslip. This "traveling phase" conceptually resembles the minuscule displacement measured between two arms in a Michelson interferometer, making iSCAT theoretically well-suited for tracking in the axial direction (and hence in 3D). In practice, however, it leads to severe instability in the iPSF signal near the microscope focal plane. The signal undergoes approximately sinusoidal contrast inversions with a half-wavelength period as the particle moves along the axial direction. As a result, the iPSF exhibits regions of very high contrast but slow variation along $z$, interspersed with regions of nearly vanishing contrast but rapid variation in $z$ (Fig. 1b). This behavior is noticeably different from fluorescence microscopy, where the signal generally peaks in the focal plane.

The impact of the unstable iSCAT contrast on localization capability can be quantified using Fisher information theory and the Cramer-Rao lower bound (CRLB) [34-38]. The CRLB represents the fundamental best-case limit for the precision achievable in fitting a parameter, such as the particle's 3D coordinates and its scattering strength. This concept is routinely used in fluorescence microscopy to analyze the suitability of engineered PSFs for single-molecule localization [6, 11, 39]. For iSCAT in the shot noise-limited case, the CRLB can be written as [34]

$$\sigma(\gamma_j) \geq \sqrt{[J^{-1}(\gamma)]_{jj}} \qquad \text{Eq. (2)}$$

where the Fisher information matrix $J$, dependent on parameters $\gamma$ (e.g., the position $x$, $y$, $z$, and the scattering strength of the particle), is defined by:

$$J(\gamma)_{ij} = \int \frac{1}{I(r)} \frac{\partial I(r)}{\partial \gamma_i} \frac{\partial I(r)}{\gamma_j} dr. \qquad \text{Eq. (3)}$$

Here, $I(r)$ is the detected signal, in our case, the iPSF, and the integral is conducted over the detector area. Intuitively, these equations convey that the more significant the change in the observed iPSF in response to a small alteration in the particle position, the more precisely the particle can be localized along the corresponding coordinate. For the general case of a translation-invariant iPSF, optimal localizability for the transverse coordinates $x$ and $y$ is achieved when the contrast is high and tightly localized. On the other hand, the best localizability in the axial direction $z$ occurs when the iPSF is rapidly changing. The combination of these factors is unfortunate from the perspective of the CRLB, as it implies a sort of uncertainty principle between the axial and transverse coordinates. In other words, the particle cannot be localized with the highest precision in both dimensions simultaneously. In Fig. 1c, numerical results supporting this prediction are presented, calculated with the focal plane positioned 5 µm above the glass coverslip. The particle contrast and CRLB exhibit pronounced oscillations near the focal plane, with the transverse ($x$ and $y$) and axial ($z$) CRLBs exactly out of phase.

Fundamentally, the unstable iSCAT contrast near the focus stems directly from the nearly flat phase of the scattered wave near the focal plane, causing all regions of the iPSF to "blink" in unison. To address this issue, we introduce an engineered iPSF, the SP-iPSF, characterized by a spatially uniformly distributed phase at all planes, thereby stabilizing the iSCAT contrast. This is achieved by imprinting a spiral phase front, ranging from 0 to 2π about the azimuthal angle, onto the scattered wave in the pupil plane of an iSCAT microscope while leaving the reference unmodified (Fig. 1a). The spiral phase, a signature of light carrying orbital angular momentum, is well known for being topologically protected upon propagation and focusing [40-42]. Previous studies have demonstrated the spiral phase modulation of the diffracted light field in the Fourier plane, acting as a spatial filter for isotropic enhancement of the edge contrast of a sample under optical microscopy [43, 44]. In this study, we reason that the spiral phase modulation preserves the uniform phase in the image plane even as the scattering particle moves along the axial direction. As a result, the iPSF evolution in response to axial motion through the focal plane is transformed from the "blinking" of the standard iPSF into a collective "winding", from a left-handed spiral, to a two-lobed dipole-like structure, to a right-handed spiral (Fig. 1d). Notably, this transformation maintains a consistent level of visibility throughout. This rectifying effect on the CRLB smoothes out the oscillations, enabling precise localization in all three dimensions simultaneously, regardless of the axial position of the particle (Fig. 1e).

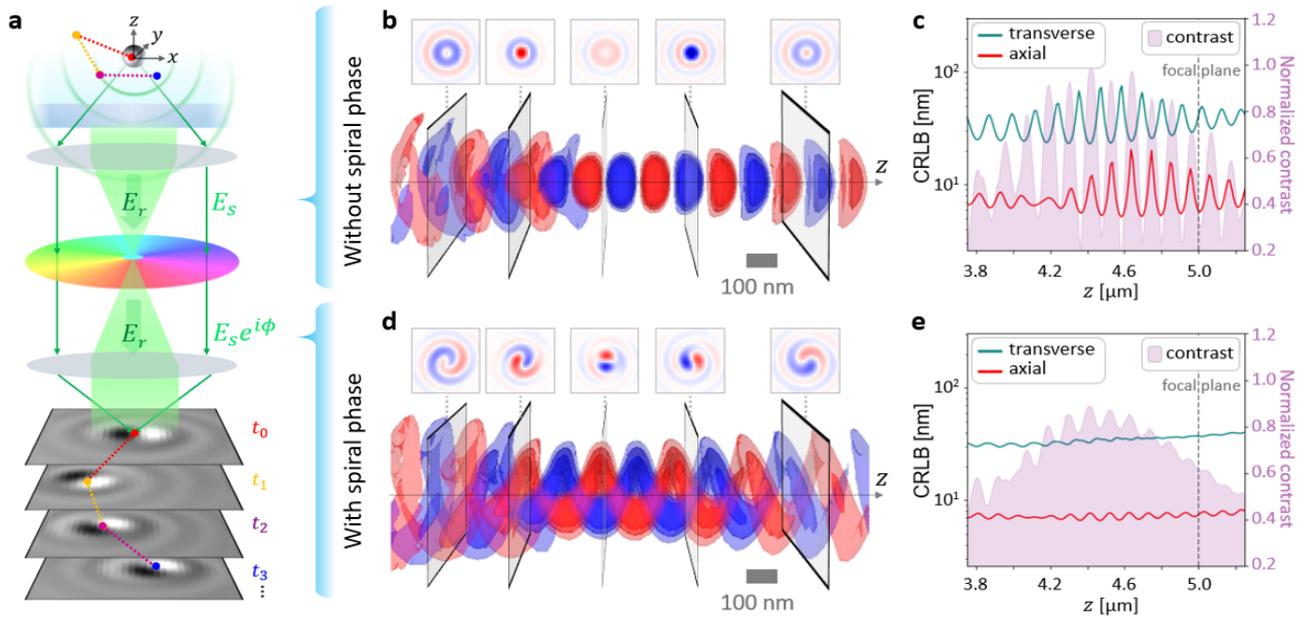

**Fig. 1. Concept of 3D single-particle localization with SP-iPSF**. **a** Spiral phase modulation of the scattered wave in the pupil plane converts the image of a nanosized scatterer into the SP-iPSF with improved localization properties. **b** Isosurfaces corresponding to contrast levels of 3, 4, and 5% for the standard particle image (iPSF) without spiral phase modulation, as a function of axial position relative to the coverslip. **c** CRLB (normalized per scattered photon) for the same case showing achievable localization in the transverse (x, y) and axial (z) directions as a function of z. The axial and transverse CRLBs are unstable and out of phase, precluding robust localization in all three dimensions. **d** The same isosurfaces as in **b**, shown for the spiral phase-modulated iPSF (SP-iPSF), where the contrast inversions are replaced by a continuous signal throughout the focus. **e** The particle contrast, and transverse and axial CRLB are smoothed out, providing reliable localizability in three dimensions at all positions.

**Robust localization performance through decoupling of interference contrast and axial position**

In addition to the fundamental localization properties described in the previous section, the SP-iPSF resolves several issues which have in the past limited the application of iSCAT to 3D SPT. The initial stage of tracking requires an accurate assessment of the initial position of the particle. In contrast to the slowly evolving PSF of darkfield or fluorescence microscopy, the oscillatory nature of the iPSF results in similar-looking profiles spaced with an approximate period of a half wavelength. This results in the existence of multiple potential solutions when attempting to fit the iPSF to a numerical model, introducing a fitting error distribution that is generally multimodal. Further complicating the situation, the standard iPSF has a symmetry about the focal plane, leading to an additional set of similarly appearing local minima on the opposite side of the focus. This increases the likelihood of confusion between motion in the positive and negative directions. The periodicity and symmetry issues are somewhat alleviated by the aberrations caused by refractive index mismatch at the coverglass/medium interface [33]. However, the differences between adjacent local minima arise mostly in the weaker, secondary ring structure about the central spot of the iPSF. These differences are difficult to resolve, especially in cases of limited SNR or when cropping to a small window—a step often necessary for PSF fitting. Consequently, the presence of multiple solutions and sign ambiguity presents significant challenges for 3D localization and tracking in iSCAT. The SP-iPSF improves upon the standard iPSF for this task by improving the distinguishability between iPSFs occurring at local intervals of approximately a half-wavelength, and by thoroughly breaking the symmetry between the positive and negative directions. Both improvements stem from the intricate evolution of the spiral iPSF. While closest to the focal plane, the particle adopts a two-lobed structure, seamlessly transitioning into a left-handed or right-handed spiral as the particle moves downwards or upwards, respectively. The spiral arms extend smoothly from the central portion of the iPSF, leading to increased visibility and distinguishability even in the case of cropping and limited SNR.

After the initial position is identified, the next step in tracking is to fit and link the particle's positions at subsequent frames to form a trajectory. When processing large numbers of particles per frame over many frames, a full global search of the parameter space for each frame quickly becomes computationally infeasible. A useful, often-used alternative is local optimization, which explores the parameter space around some initial starting guess. This is well-suited to high-speed 3D-SPT, where the fitting output of the current frame can be used to initialize the fitting parameters for the next one. Specifically for the axial positioning, as long as the particle does not travel too long of a distance between frames, a local optimization based on least-squares minimization or other optimization routines should converge to the correct solution. However, fitting to the standard iPSF is complicated by a high degree of coupling between the scattering strength and the axial position—both are reflected to a large degree in the iSCAT contrast. In the general case where the particle scattering amplitude is not known a priori, this coupling inevitably leads to errors in the axial localization.

These concepts can be visualized by plotting the mean-squared error as a two-dimensional function of axial position and particle scattering amplitude, while for simplicity holding the transverse position fixed. On this error surface, each point represents a potential iPSF fit, and the value at that point reflects the similarity of said fit to the data. The process of fitting a model iPSF to the data then becomes the problem of finding the global minimum of this error surface. For the standard iPSF, the coupling between the two parameters causes each potential solution on the error surface to take on a characteristic U-shaped shape rather than a well-formed minimum (Fig. 2c). As a result, in the presence of noise, or slight differences in the experimental iPSF from the fitting model, even small variations in the initial parameter guesses can lead to incorrect results for the fitting. In contrast, the uniform phase distribution of the SP-iPSF has the effect of completely decoupling the scattering strength and axial position, such that the former is represented in the iPSF contrast, while the latter is encoded instead in its shape. This has the effect of reshaping the error landscape into one which can be more robustly fit (Fig. 2d).

We carried out numerical simulations to demonstrate this effect. We calculated the iSCAT signal for an 80 nm silica particle positioned at $z = 4.615$ um above the glass substrate, close to the "effective focus" where the particle scattering signal is the strongest for both the standard iPSF (Fig. 2a) and SP-iPSF (Fig. 2b), and created simulated data by corrupting the clean signal (top) with realistic photon shot noise (bottom). We generated 1,000 of these noisy frames for each iPSF, and fit them independently to the model as a function of amplitude and axial position (with the transverse position fixed at the origin). The fitting was carried out in two steps: first, by doing an approximate global fit to the axial position of the particle by comparison to a dictionary of normalized iPSFs at coarsely sampled $z$, and second, by refining the contrast and $z$ through an iterative least-squares minimization of the mean squared error.

The results of the fitting are shown in Fig. 2c-f as yellow circles overlaid on the error surface, with the true solution indicated by the intersection of the dashed orange lines. For the standard iPSF, similar depths between different U-shaped local minima make the initial estimate of the particle position difficult, leading to the possibility of large fitting errors. Furthermore, the U-shape of each individual minimum leads to increased uncertainty and coupled errors in both the axial position and scattering amplitude (Fig. 2d) [45]. In contrast, for the SP-iPSF, the reshaping of the error surface is immediately reflected in improved fitting results, as in each case the correct minimum is identified, and the precision in both position and amplitude increased and decoupled. We emphasize that both the simulated data and candidate fits are generated from the same numerical model, so the results shown reflect the inherent differences between the two iPSFs.

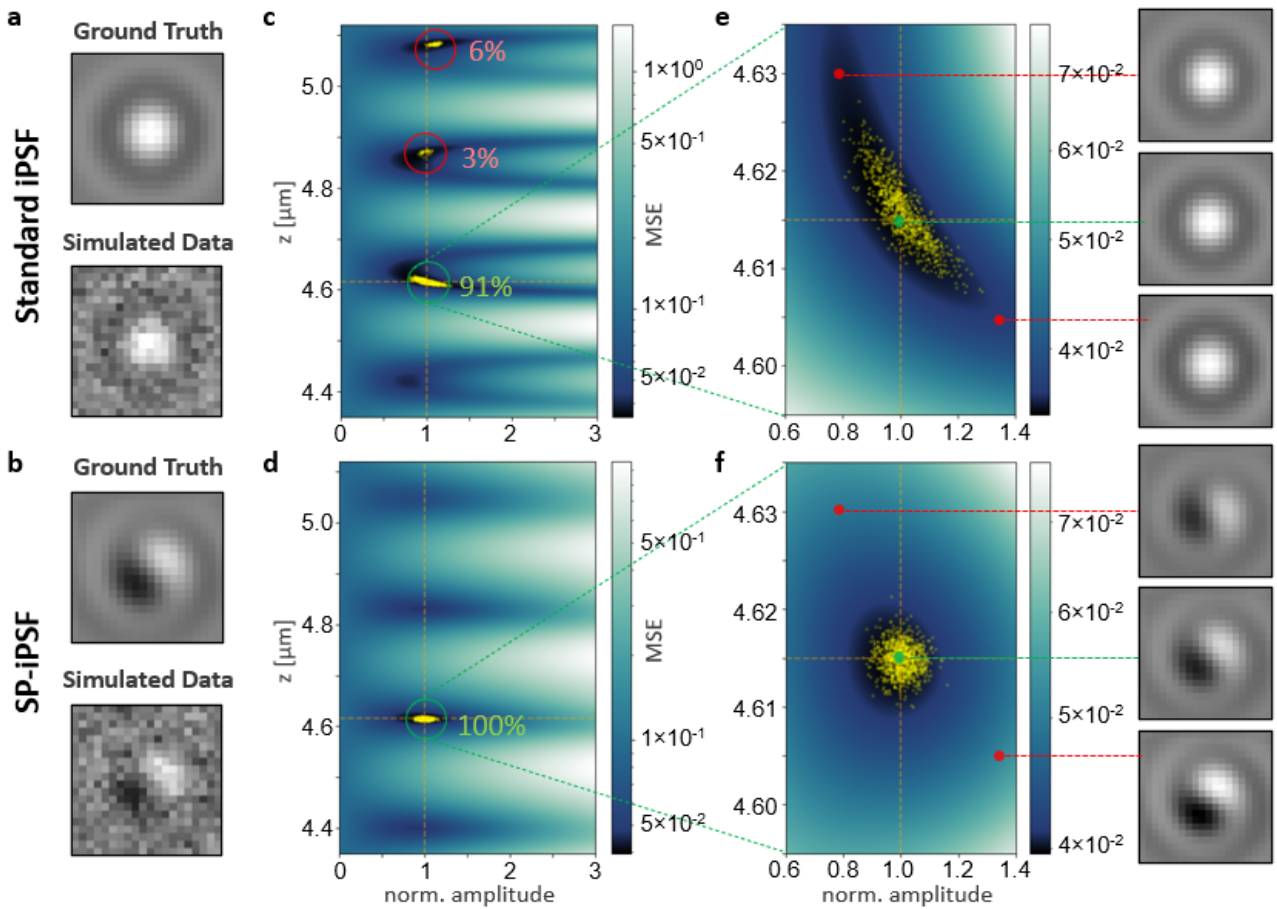

**Fig. 2. Simulation showing increased fitting robustness with SP-iPSF through symmetry breaking and z/contrast decoupling.** iSCAT images for an 80 nm silica particle centered on the optical axis and at the z = 4.615 um above the substrate are simulated for both **a** the standard iPSF and **b** SP-iPSF, corrupted with realistic noise to generate 1000 frames, and fit independently. **c** At certain z-positions, high similarity between adjacent local minima with the standard iPSF lead to the possibility of severe fitting errors. **d** The SP-iPSF reduces this similarity, decreasing the probability that a global fit will end up in the wrong local minimum. **e** Within a specific local minimum, the standard iPSF has a U-shaped trough-like error landscape to coupled errors and reduced precision in both z-position and contrast. **f** With the SP-iPSF, the error surface is reshaped into a well-formed local minimum due to decoupling of z-position and contrast, leading to increased fitting precision in both.

**Demonstration of high-speed 3D particle tracking with SP-iPSF**

We demonstrate the benefits of the SP-iPSF for SPT both numerically and experimentally. We simulate a 3D random walk based on the diffusion equation for an 80 nm diameter particle diffusing in water (Fig. 3a). Based on this trajectory, we generate simulated iSCAT images based on typical experimental parameters (wavelength = 532 nm, NA = 1.33, pixel size = 72 nm, focal plane 5 μm above glass, particle refractive index = $n_{silica}$ = 1.4585). To account for the effects of particle motion during the camera exposure, we use our iPSF model to generate simulated images at a factor of 3 above the final camera frame rate and bin them together to create realistic motion-blurred data at a frame rate of 20,000 Hz. Finally, we corrupt the images with Poisson (shot) noise to reflect typical experimental SNR. The simulation is repeated for both the standard iPSF and SP-iPSF cases, while keeping the trajectory, SNR, and all other parameters fixed.

We carry out 3D SPT on the simulated images by doing localizations on each frame by least squares minimization with respect to four parameters: $x$, $y$, $z$, and scattering amplitude. Note that the scattering amplitude shown here is normalized to that of an 80 nm silica particle based on a noise-free simulation. The $x$ and $y$ positions of the first frame are given as input and used to crop the particle image to a 20x20 pixel window, and the initial z-position is estimated by a coarse local optimization as in the previous section. Cropping and fitting for all subsequent frames are then initialized based on the result of the most recent successful localization.

The results are shown for the standard iPSF (Fig. 3b) and SP-iPSF (Fig. 3c). At the simulated SNR, both the regular iPSF and SP-iPSF are able to correctly fit the trajectory with nanometric precision, while also estimating the scattering amplitude. However, the SP-iPSF has significantly reduced localization error for all four parameters. The most noticeable improvements are in the axial dimension and amplitude, a direct result of the decoupling described in the previous section. The improvement in the transverse dimensions is a result of the stabilization of the SNR throughout the focal plane, such that at positions where the standard iPSF signal nearly vanishes (Fig. 3d) the SP-iPSF still presents a strong signal (Fig. 3e). This difference becomes far more significant in the presence of lower SNR, or non-uniform background as is generally present in a real experiment. In such cases, the vanishing contrast of the standard iPSF can cause the fitting to completely fail, resulting in an incorrect or broken trajectory.

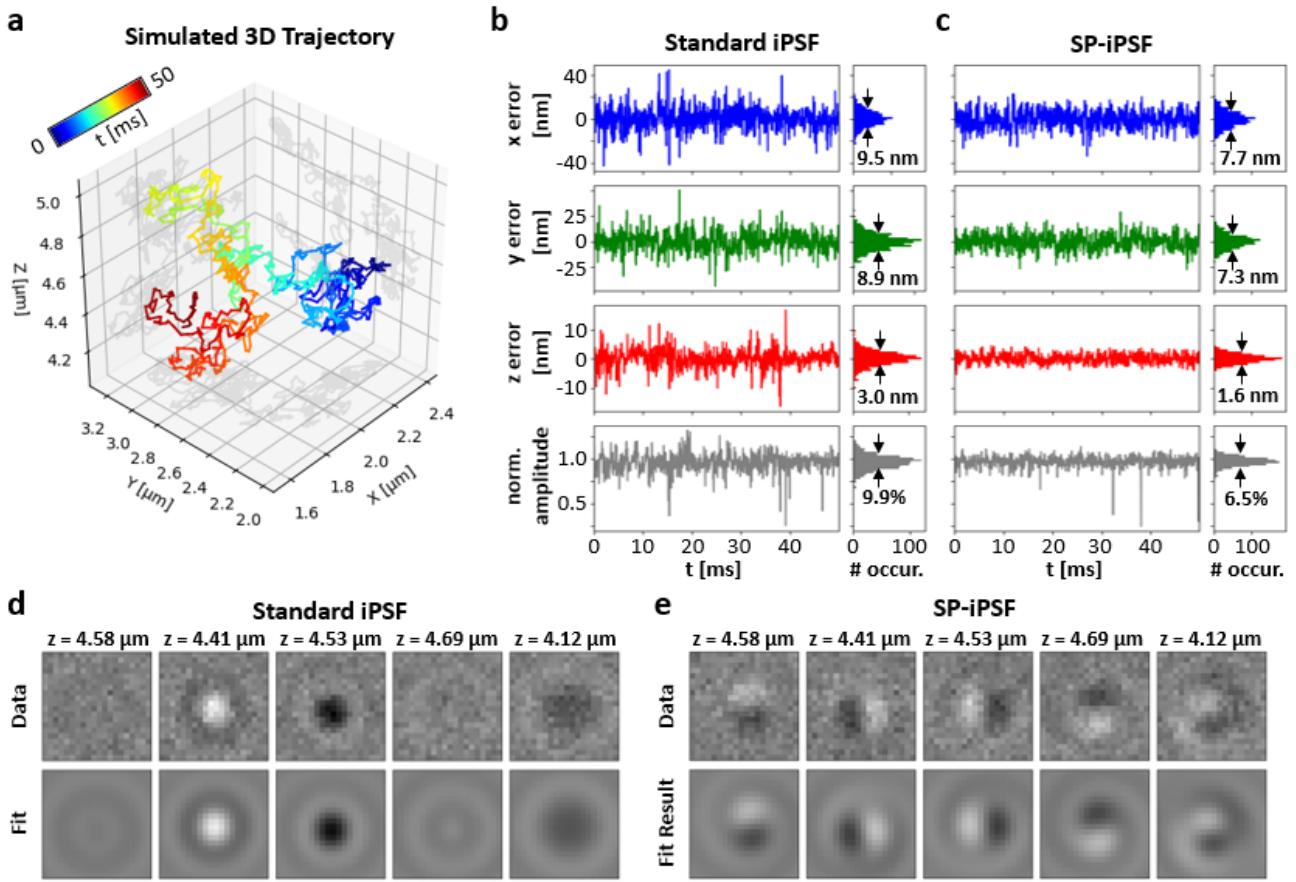

**Fig. 3: Numerical demonstration of 3D nanoparticle tracking with SP-iPSF**. **a** Simulated trajectory for an 80 nm silica particle diffusing in water, sampled at a rate of 20 kHz. Realistic iSCAT simulations based on this trajectory are fit to a 3D iPSF model. Fitting error results for the **b** standard iPSF, and **c** SP-iPSF show the superior fitting precision of the SP-iPSF in all three dimensions, as well as scattering amplitude assessment. Sample cropped data frames (top) and fits (bottom) for **d** standard iPSF show inconsistent SNR at different axial positions near the focal plane, while the **e** SP-iPSF has consistent SNR and instead reflects the axial position by a change in shape and orientation.

We next verify the 3D SPT with an experimental demonstration, tracking the motion of a nominally 80 nm diameter silica nanoparticle diffusing in water at a frame rate of 20,000 Hz. We experimentally implement the SP-iPSF by augmenting an iSCAT microscope with a 4f-system after the sample, and placing a reflective spatial light modulator (SLM) in the Fourier plane. On the SLM, the reference and scattered waves are well separated in spatial frequency. We apply a phase mask where the central, low-frequency portion has a flat phase for the reference beam, while the higher spatial frequencies have the appropriate spiral phase (Fig. 1a). The sample is prepared by creating a flow chamber composed of nanoparticles in water sandwiched between two glass coverslips. The experimental parameters such as wavelength and NA are similar to those in the simulation.

We remove the non-uniform background noise by image postprocessing and identify and crop candidate particles based on a machine-learning-assisted model. Then, the same localization and tracking routines are applied to the experimental data. The results are shown in Fig. 4. The reconstructed trajectory (Fig. 4a) samples the particle diffusing within a volume of ~1 μm³ over the time course of ~65 ms. The diffusive motion behavior in all three dimensions is visually consistent with that expected of a freely diffusing particle, and the measured scattering amplitude remains approximately constant even as the particle moves through a range of axial positions (Fig. 4b). Since we do not have a ground truth for the particle diffusion trajectory, we analyze the standard error returned by the fit for all four fit parameters (Fig. 4c). We find that the localization precision in all three dimensions is on the nanometer scale. Notably, the precision in the axial direction is significantly higher than that in the transverse directions, due to the interferometric sensitivity—consistent with our theoretical and numerical results. Sample cropped data frames and the corresponding best-fit images are shown in Fig. 4d, taken from throughout the trajectory and sorted by axial position. These images illustrate the sensitivity of SP-iPSF to nanoscopic axial displacements (compare z = 4.35, 4.37, and 4.53 μm), and the subtle but clear differences that distinguish the periodic local minima (z = 4.53 μm and 4.76 μm).

In addition to the 3D particle localization, a highly accurate and precise scattering amplitude is measured with an error of only 4.41%, thanks to the decoupling of scattering amplitude and axial position in SP-iPSF. The mean amplitude over the trajectory length is ~28% larger than the expected value for a nominally 80 nm silica particle (Fig. 4b, gray), consistent with a particle of slightly larger diameter 87 nm. To check this, we make a complementary measurement of the particle size based on the 3D diffusion trajectory, calculating 3D mean-square displacements (MSD) as a function of a time interval for each trajectory (Fig. 4e) [46],

$$MSD(n\Delta t) = \frac{1}{N-n}\sum_{i=1}^{N-n}\{\mathbf{r}[(i+n)\Delta t] - \mathbf{r}(i\Delta t)\}^2. \qquad \text{Eq. (4)}$$

where $\mathbf{r}(t)$ is the particle 3D position at time $t$, $\Delta t$ is the frame time, and $N$ is the trajectory length. The MSD of the trajectory exhibits a linear dependency on the time interval, indicating free Brownian motion as expected for our nanoparticle colloids. For the Brownian particle, the diffusion coefficient (D) can be estimated from the slope of the MSD [47]:

$$MSD(n\Delta t) = 6D(n\Delta t) + \sigma_e^2. \qquad \text{Eq. (5)}$$

Here, $\sigma_e$ is a constant offset due to the localization uncertainty. According to the single-particle MSD analysis, we determine D = 5.12 μm²/s for the particle. We can then estimate the hydrodynamic diameter of the particle based on the measured D using Stokes-Einstein equation. Given the viscosity of water as 0.9775 mPa·s at a temperature of 21 degrees C, the hydrodynamic diameter of this specific silica nanoparticle is estimated as 86 nm, in agreement with the result obtained by the amplitude measurement.

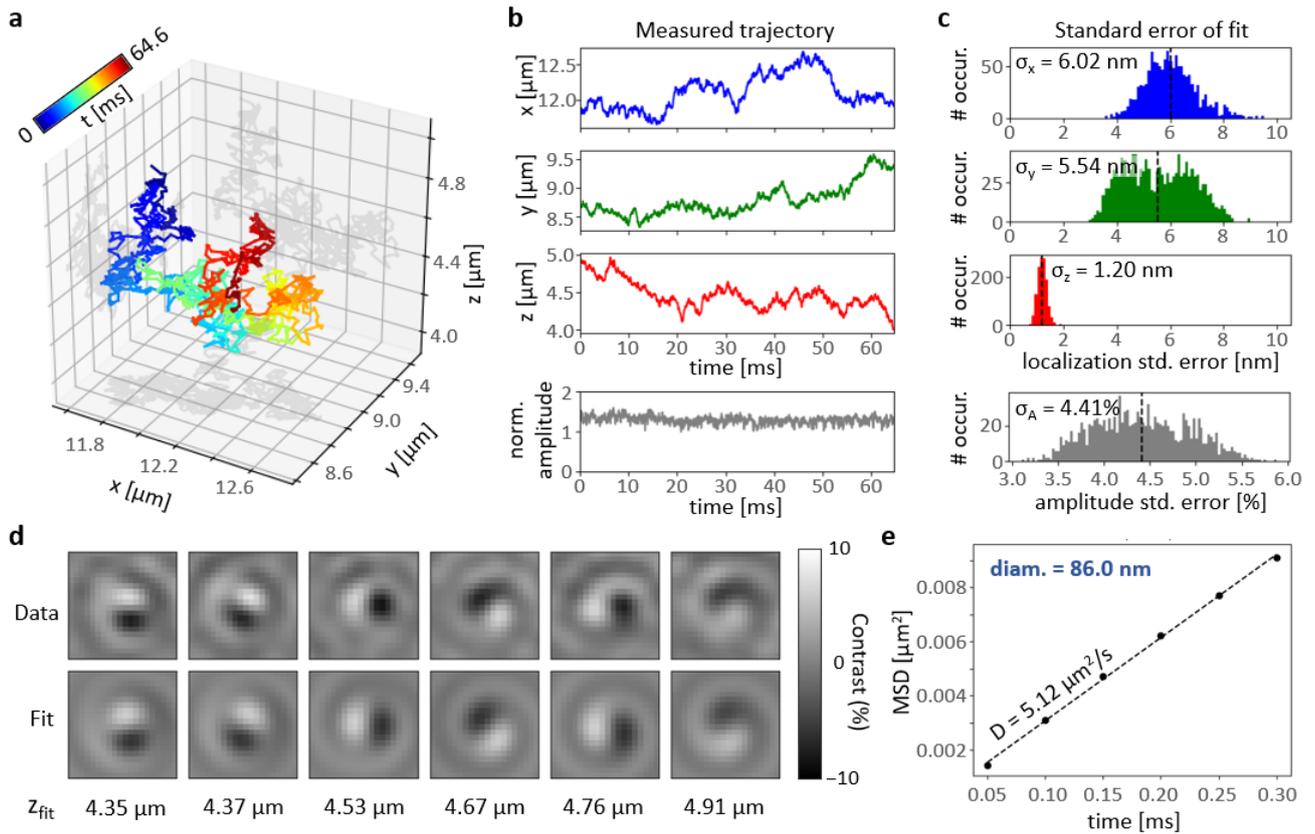

**Fig. 4. Experimental demonstration of 3D tracking of a single nanoparticle**. **a** Sample 3-dimensional trajectory for a nominally 80 nm diameter silica sphere undergoing free diffusion in water at 20 kHz over 64.6 ms (1,292 frames). **b** Fitting results for the x, y, and z positions as a function of time, as well as the particle scattering amplitude. **c** Histograms of estimated errors in x, y, and z positions and scattering amplitude. **d** Experimental particle images at representative heights and their corresponding fits, showing excellent agreements throughout the depth of focus. **e** Linear fit to the first few points of the MSD curve produces a measurement of the particle diffusion coefficient and the hydrodynamic diameter of the particle. Both the scattering amplitude and diffusion behavior are consistent with a silica nanoparticle with a diameter of approximately 86 nm, slightly larger than the nominal value.

**Characterization of polydisperse nanoparticle samples by 3D particle tracking with SP-iPSF**

To show an example application enabled by SP-iPSF, we demonstrate simultaneous characterization of particle size and electric dipole polarizability ($\alpha_{dip}$) with a polydisperse nanoparticle mixture. The polydisperse nanoparticle mixture consists of three different particle samples, i.e., 80 nm silica, 100 nm polystyrene, and 120 nm silica nanoparticles in water. The particle size is estimated based on the MSD analysis as described earlier. The hydrodynamic diameters of these three particle samples are measured as 85.5±15 nm, 99.8±8 nm, and 114±13 nm, respectively, agreeing well with the manufacturer specifications with ~10% uncertainty.

In addition to the diffusion trajectories, we determine the $\alpha_{dip}$ of individual particles from the measured scattering amplitudes based on the numerical simulation with a dipole approximation.

$$\alpha_{dip} = 4\pi a^3 \frac{\varepsilon_p - \varepsilon_m}{\varepsilon_p + 2\varepsilon_m}. \qquad \text{Eq. (6)}$$

here $a$ is the particle radius, $\varepsilon_p$ and $\varepsilon_m$ are the permittivities of the particle and the surrounding medium, respectively. Quantitative measurement of $\alpha_{dip}$ allows us to distinguish nanoparticles of different materials even if they are of similar sizes. We verify that the SP-iPSF fitting results can be used to make continuous assessments of the particle $\alpha_{dip}$ over a range of axial positions. Considering the fitting results over the full trajectory, we determine the $\alpha_{dip}$ of individual particles with a high precision.

The combination of the $\alpha_{dip}$ and the hydrodynamic size of a particle deduced from the 3D SPT *via* SP-iPSF allows for accurate characterization of nanoparticles. Fig. 5a plots the relationship between the $\alpha_{dip}$ and the particle size for the three different particles. These three different particle species are clearly distinguished, illustrating the high accuracy and precision of our particle tracking analysis. We stress that, compared to the existing SPT methods with a standard PSF, the unique opportunity offered by SP-iPSF is the accurate measurement of $\alpha_{dip}$ through 3D continuous characterization. For small Rayleigh particles of the same refractive index, a linear dependency is expected between the $\sqrt[3]{\alpha_{dip}}$ and the particle diameter. Indeed, the data of 80 and 120 nm silica nanoparticles fall on a line with a constant refractive index of 1.4607 that matches with that of silica [48]. On the other hand, 100 nm polystyrene nanoparticle data fall on a different line that corresponds to the refractive index of 1.5983 of polystyrene [49].

Fig. 5b displays representative 3D diffusion trajectories of the three particle species. The MSD curves of the three particle species are plotted in Fig. 5c, all of which exhibit linear dependency on the time interval that agrees well with the model of free diffusion. Taken together, the SP-iPSF facilitates accurate particle tracking and scattering amplitude quantification for freely diffusing single nanoparticles. The combination of these capabilities offers the opportunity to characterize complex nanoparticle samples with heterogeneous sizes and compositions at single-particle resolution.

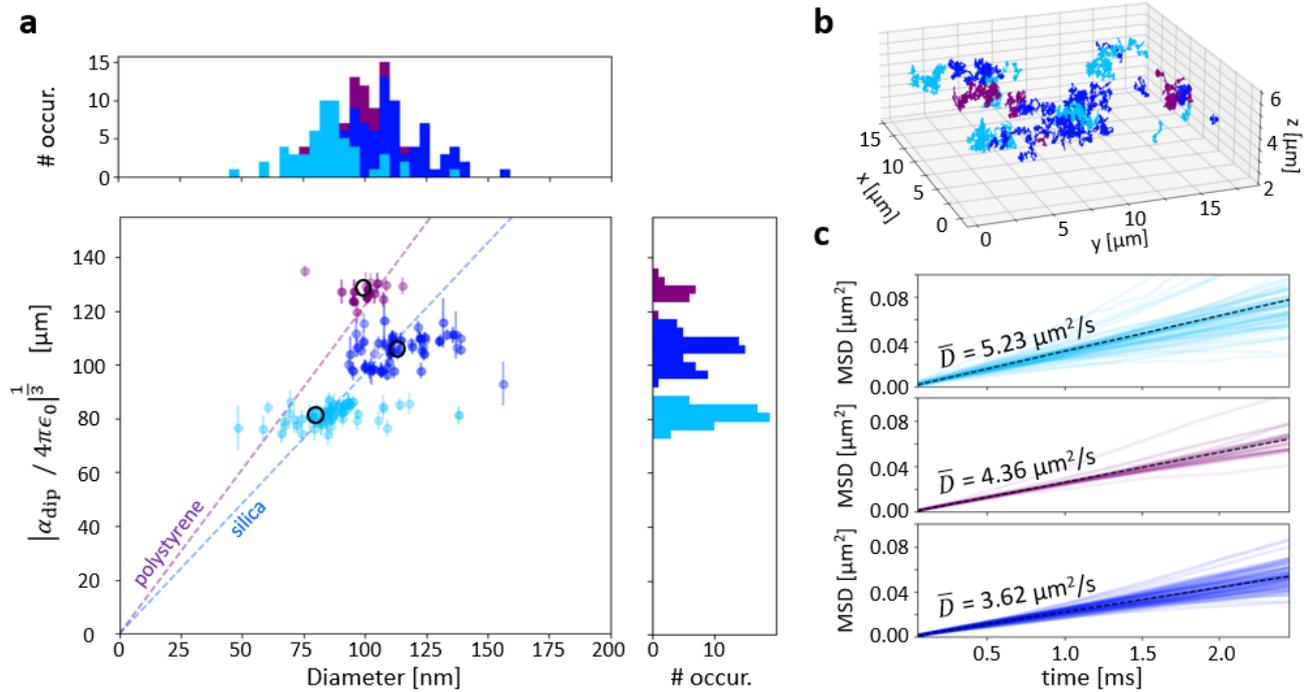

**Fig. 5: Application of SP-iPSF based 3D SPT for optical particle sizing and refractive index measurement. a** Scattering amplitude and hydrodynamic diameters are measured from 132 individual trajectories from a polydisperse solution of nominally 80 nm silica, 120 nm silica, and 100 nm polystyrene spheres. The transparency of each point is inversely proportional to the square root of the trajectory length, and the amplitude error bars taken from the standard deviation of the fits over the trajectory. The dashed lines indicate the theoretical relationship between amplitude and diameter in the Rayleigh scattering approximation given literature values for the refractive index of silica and polystyrene. Color indicates classification based on a 2D Gaussian mixture model, and the black circles indicate the mean polarizability and diameter values for each class of particle. **b** Measured trajectories plotted in the full 3D volume of observation, colored as in **a**. **c** The first 50 points of the MSD curves for all individual trajectories (colored, transparency weighted by length as in **a**) separated by the same classification, and linear fits (black line) to the ensemble average MSD of all trajectories. The calculated hydrodynamic diameters are in good agreement with the nominal values.

**DISCUSSION**

iSCAT microscopy has the potential to localize very small particles in 3D due to its remarkable sensitivity and high axial sensitivity. However, due to the characteristics of the standard iPSF, it has not been demonstrated to enable robust 3D tracking of weakly scattering particles. In this study, we introduced iPSF engineering, which mitigates the problems associated with the regular iPSF while retaining the high axial sensitivity, thus unlocking the power of iSCAT for robust 3D particle tracking. The proposed SP-iPSF encodes the axial position of the particle in the radial phase, which is detected through interferometry in the measured intensity. As the SP-iPSF provides persistent signals in the axial direction, it enhances the robustness and accuracy in particle localization when compared to the regular iPSF. The improved iSCAT 3D SPT has immediate applications, including measuring dynamic movements of virus particles [29, 50, 51], cell vesicles [31, 52], and nano-sized optical probes to single molecules [30, 53], where the 3D displacements are crucial in understanding their interactions with complex environments. Accurate 3D SPT also helps improve the performance of nanoparticle characterizations by nanoparticle tracking analysis (NTA) where the hydrodynamic size of a particle is estimated based on the free diffusion trajectory [2, 54]. It is worth noting that, in contrast to non-interferometric approaches, SP-iPSF allows for more accurate tracking in the axial direction than in the lateral direction.

Another powerful capability provided by SP-iPSF is accurately quantifying the magnitude of the scattered field of a particle in every image, even when the particle is not perfectly located at the focal plane. Multiple observations of the scattering amplitude of a particle are possible when it diffuses within the detection volume. This enables continuous monitoring of the scattered field of a freely diffusing particle, making it possible to capture time-evolving particle mass. If the particle mass does not change over time, the accuracy in determining the particle scattering amplitude can be significantly improved by averaging the results of these multiple observations. The ability to accurately measure the scattering amplitude of a freely diffusing particle eliminates the requirement of placing the particle at the focal plane under iSCAT for characterization, which is generally achieved by depositing the particle on a substrate with stable sample positioning [55, 56]. When no supporting substrate is needed, the background scattering of the substrate roughness will no longer be an issue when measuring the weakly scattering particles [57].

This study reports proof-of-principle iPSF engineering for enhancing iSCAT SPT, and several improvements can be made to further improve the performance. Replacing the SLM with a phase plate is expected to reduce the polarization dependency and improve the phase stability. In addition, the particle contrast can be enhanced by attenuating the reference beam in the Fourier plane [58], allowing stronger illumination for the detection of smaller particles or even single macromolecules. Regarding the localization algorithm, we used simple least squares minimization for 3D particle localization of the SP-iPSF in the current study. We anticipate that machine learning-assisted algorithms can potentially realize faster localization and be more immune from background noise [59] . Finally, in analogy to fluorescence PSF engineering for 3D SPT, iPSF engineering using different phase patterns should make it possible to enhance the dynamic range for SPT at different signal-to-background ratios. We anticipate that the integration of iPSF engineering into iSCAT will empower single-particle measurements with unparalleled sensitivity and accuracy. This breakthrough holds the potential to unveil new opportunities for exploring nanoscale sciences with unprecedented clarity.

**Funding**

This work is supported by the iMATE program, Academia Sinica, Taiwan (AS-iMATE-111-35); National Science and Technology Council (NSTC), Taiwan (NSTC 111-2112-M-001-051-MY5). We thank to National Center for High-performance Computing (NCHC) of National Applied Research Laboratories (NARLabs) in Taiwan for providing computational and storage resources.

**Acknowledgment**

The authors thank Huan-Hsin Tseng and Yu Tsao for useful discussion on deep learning-assisted algorithms for nanoparticle detection. N. J. B. thanks Academia Sinica for the Postdoctoral Scholarship.